\documentstyle[sprocl]{article}

\input{psfig}

\def\be{\begin{equation}}
\def\ee{\end{equation}}
\def\bea{\begin{eqnarray}}
\def\eea{\end{eqnarray}}

\begin{document}

\title{TRANSVERSE MOMENTUM BROADENING DUE TO THE MULTIPLE SCATTERING}

\author{XIAOFENG GUO}

\address{Department of Physics, Columbia University, New York, 
NY 10027, USA} 


\maketitle\abstracts{ 
Using the Drell-Yan process in hadron-nucleus collisions
and deeply inelastic lepton-nucleus scattering (DIS) 
as examples, I show that the transverse momentum 
broadening can be expressed in terms of four-parton correlation 
functions. I argue that jet broadening in DIS provide  
an excellent measurement of the four-parton correlation functions and  
a test of QCD treatment of the multiple scattering.}

\section{Introduction}

Most high energy collisions can be described by a single hard scattering.
By studying the single scattering, we can extract the  parton distribution 
functions and test the short-distance dynamics of strong interactions.  
The normal parton distributions have 
the interpretation of the probability distributions to find a parton
within a hadron.  On the 
other hand, in order to study quantum correlations of multi-partons 
inside a hadron, we need to investigate the multiple scattering. 
The multiple scattering is 
a high twist effect and is small compare to the single scattering. 
However, in collisions involve nucleus, the nucleus environment 
will enhance the multiple scattering effect. For example, 
when a parton propagates through a nuclear matter, the average 
transverse momentum is broadened due to the multiple scattering. 
Nuclear dependence of such broadening provides an 
excellent probe to study QCD dynamics beyond the single 
hard-scattering picture. 

A reliable calculation of the multiple scattering in QCD 
perturbation theory 
requires to extend the factorization theorem \cite{Factorization} 
beyond the leading power.  Qiu and Sterman showed that the factorization 
theorem for hadron-hadron scattering holds at 
the first-nonleading power in momentum transfer \cite{QS}, 
which is enough to study double scattering processes in hadronic 
collisions. Applying  this generalized factorization theorem, 
Luo, Qiu and Sterman (LQS) developed a consistent treatment of 
the multiple scattering at the partonic level \cite{LQS1}.
LQS expressed the nuclear dependence of di-jet momentum imbalance
in photo-nucleus collisions in terms of four-parton (twist-4) nuclear 
correlation functions \cite{LQS2}. 
In the following, I will use the
the  deeply inelastic lepton-nucleus scattering (DIS)
and the Drell-Yan process in hadron-nucleus collisions as examples
to show that transverse momentum broadening is proportional to the
similar four-parton correlation functions \cite{GuoDIS}.

The predictions of the multiple scattering effects rely on accurate 
information of these four-parton correlation functions.
The four-parton (twist-4) correlation functions are as fundamental as 
the normal twist-2 parton distributions. 
Just like normal parton distributions,
multi-parton correlation functions are non-perturbative and universal.  
QCD perturbation theory cannot provide the absolute prediction of these
correlation functions.  They can be measured in some
processes and be tested in other processes.  
Using the Fermilab E683 data, LQS estimated the 
size of the relevant twist-4 parton correlation functions to be of 
the order of $0.05-0.1$ GeV$^2$ times typical twist-2 parton 
distributions \cite{LQS2}.
However, Fermilab and CERN data \cite{E772,NA10} 
on nuclear enhancement of the average Drell-Yan transverse momentum,  
$\Delta \langle q_T^2 \rangle $,  prefer a much smaller size of the 
four-parton correlation functions \cite{GuoDIS,YuriD}. 
This discrepancy may result from different higher order contribution to 
nuclear enhancement of the average Drell-Yan transverse momentum 
\cite{Guo,GQS} 
and the di-jet momentum
imbalance. It is necessary to 
study the higher order corrections to these two observables 
in order to test QCD treatment of multiple scattering. 
Meanwhile, it is also important to use jet broadening in 
DIS as an independent measurement of the four-parton correlation 
functions \cite{GuoDIS}.

In the following, I first present the derivation of  
the transverse momentum broadening in Drell-Yan process and 
jet broadening in DIS. I then argue that jet broadening in DIS is 
an excellent observable to study the four-parton correlation 
function. 

\section{Transverse momentum broadening for the Drell-Yan pairs}

Consider the Drell-Yan process in hadron-nucleus collisions, 
$  h(p') + A(p) \rightarrow \ell^+\ell^-(q) + X $, where 
$q$ is the four-momentum for the virtual photon $\gamma^*$ 
which decays into the lepton pair.  $p'$ is the momentum for 
the incoming hadron and $p$ is the momentum per nucleon 
for the nucleus with the atomic number $A$. 
In order to extract the effect due to the multiple scattering, 
we define the Drell-Yan transverse momentum broadening as 
\begin{equation}
\Delta \langle q_T^2\rangle 
\equiv \langle q_T^2 \rangle ^{hA}
      -\langle q_T^2 \rangle ^{hN} \ ,
\label{dydqt2}
\end{equation}
with $q_T$  the transverse momentum of the Drell-Yan pair, and the
the averaged transverse momentum square is defined as  
\begin{equation}
\langle q_T^2\rangle ^{hA}=
\left. \int dq_T^2 \cdot q_T^2 \cdot
\frac{d\sigma_{hA}}{dQ^2dq_T^2} \, \right/ 
\frac{d\sigma_{hA}}{dQ^2} \ .
\label{qt2}
\end{equation} 
In Eq.~(\ref{qt2}), $Q$ is the total invariant mass of the lepton 
pair with $Q^2=q^2$.
From our definition, $\Delta \langle q_T^2\rangle$ represents a 
measurement of QCD dynamics beyond the traditional single-hard 
scattering picture.  The nuclear dependence of 
$\Delta \langle q_T^2\rangle$, defined 
in Eq.~(\ref{dydqt2}), is a 
result of the multiple scattering between the parton from incoming  
beam and the nuclear matter before the Drell-Yan pair are produced.

\begin{figure}[hbt]
\begin{minipage}[t]{2.0in}
\psfig{figure=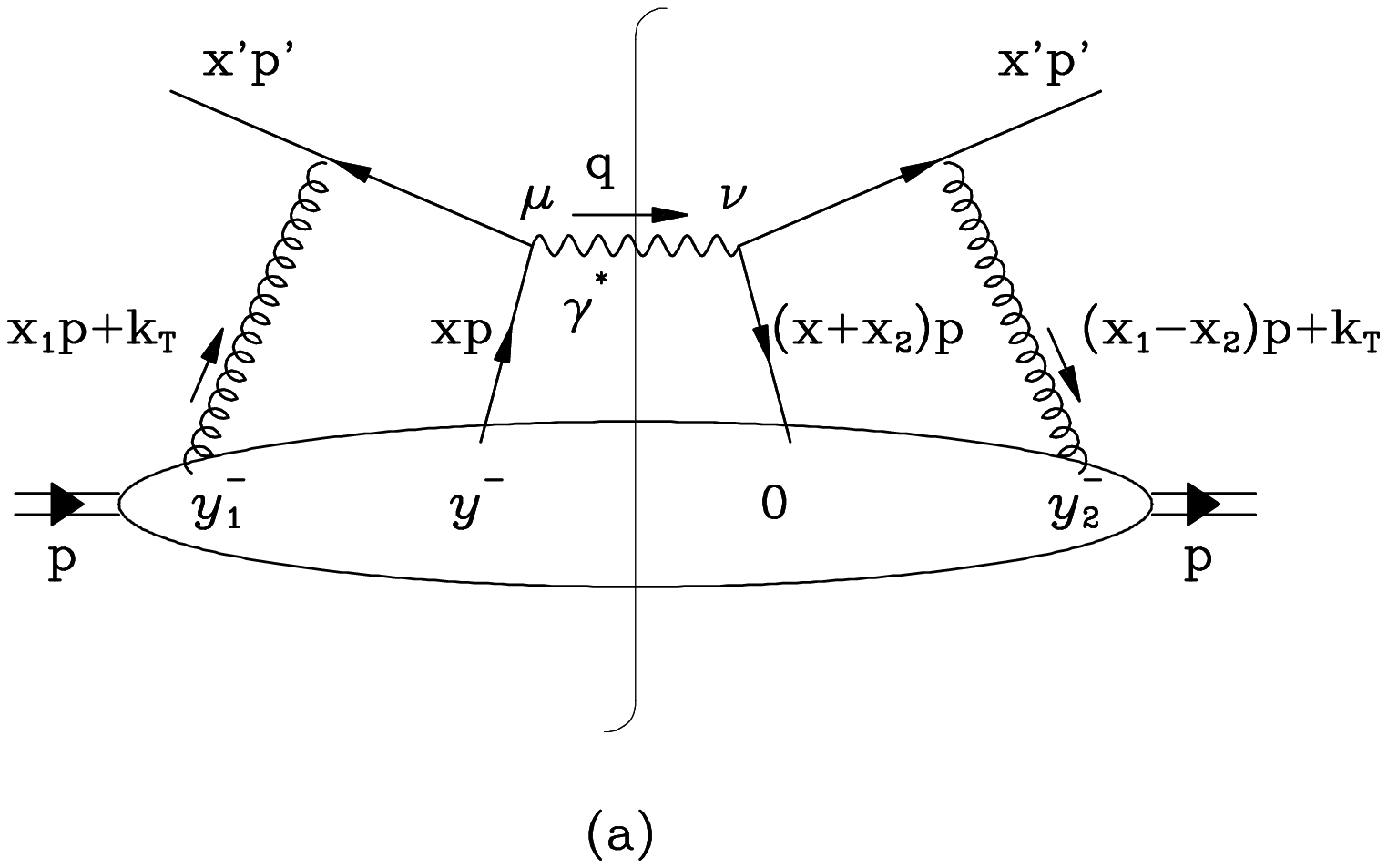,width=1.0in}
\end{minipage}
\hfill
\begin{minipage}[t]{2.0in}
\psfig{figure=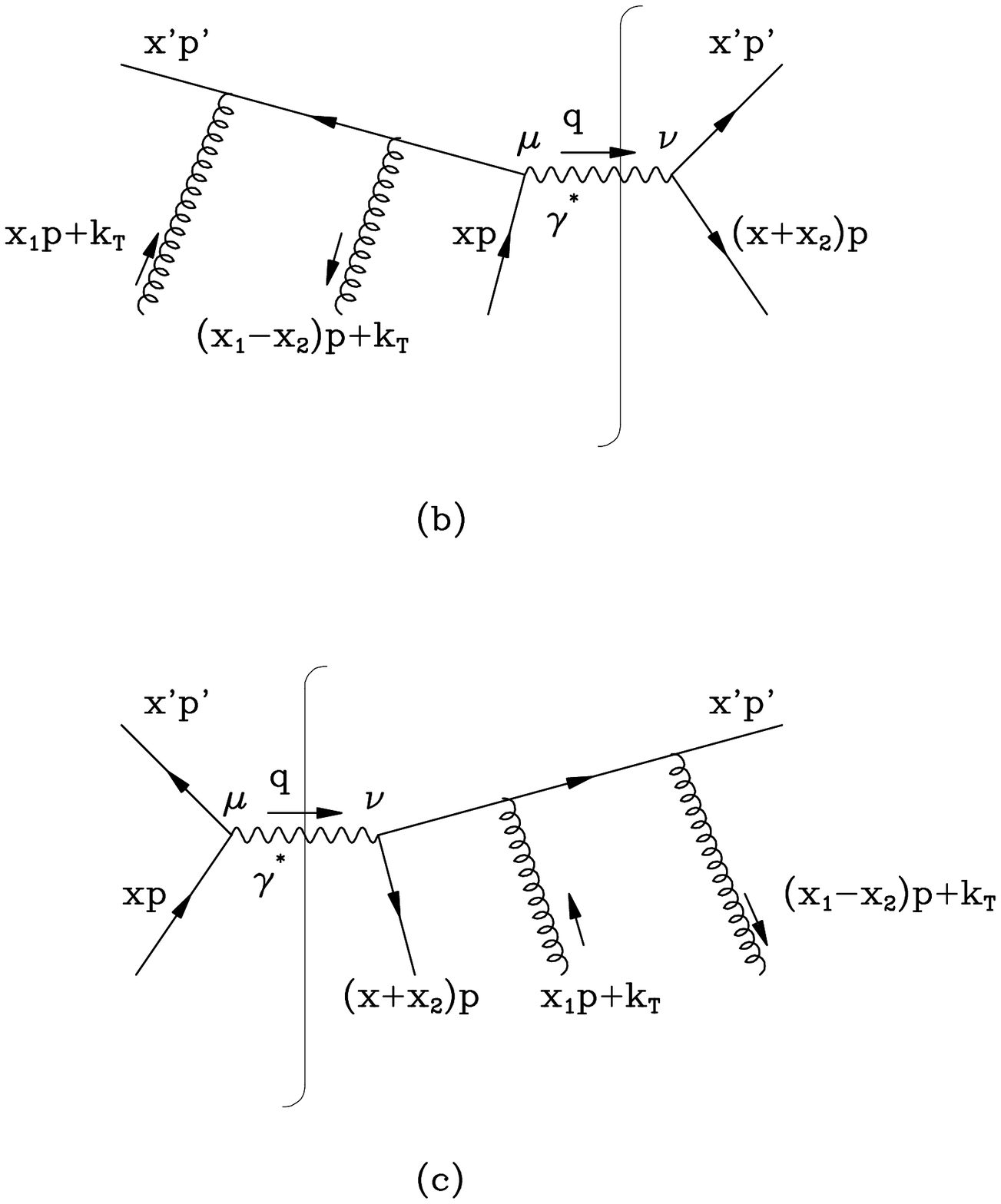,width=1.0in}
\end{minipage}
\caption{Lowest order double scattering contribution to 
the nuclear enhancement of Drell-Yan $\langle q_T^2 \rangle$; 
(a)symmetric diagram; (b) and (c): interference diagrams.} 
\label{fig1}
\end{figure}

If we keep only the double scattering contribution 
and neglect contribution from 
the higher multiple scattering, we have 
\begin{equation}
\Delta \langle q_T^2\rangle 
\approx 
\int d q_T^2 \cdot q_T^2\cdot 
\frac{d\sigma_{hA}^D}{dQ^2 dq_T^2}
\,    \left/ \frac{d\sigma_{hA}}{dQ^2} \right.  \ ,
\label{dydqt2a}
\end{equation} 
where superscript ``D'' indicates the double scattering 
contribution.  Fig.~\ref{fig1} shows the leading order double scattering 
between the parton ``f'' from incoming beam and the nucleus.
According to the generalized factorization theorem \cite{QS},
the double scattering cross section can be expressed as
\begin{equation}
d\sigma^D_{hA \rightarrow \ell^+ \ell^-}
= \left(\frac{2 \alpha_{em}}{3Q^2} \right) 
\, \sum_f\, \int dx'\, \phi_{f/h}(x') \cdot 
d\hat{\sigma}^{D}_{fA\rightarrow\gamma^*}(x',q)\ ,
\label{dydsigma}
\end{equation}
with $d\hat{\sigma}_{fA}^D$
the parton level double scattering cross section, and
\begin{eqnarray}
d\hat{\sigma}^D_{fA} &=& \frac{1}{2x's} \,
\int dx\, dx_1\, dx_2 \int d k_T^2 \, \bar{T}^{(I)}(x,x_1,x_2,k_T)\, 
\nonumber \\
&\ & \times 
\bar{H}(x,x_1,x_2,k_T,x'p',p,q) \ .
\label{dyD}
\end{eqnarray}
In Eq.~(\ref{dydsigma}), $\phi_{f/h}(x')$ is the parton distribution 
from the hadron.
In Eq.~(\ref{dyD}), $\bar{H}$ is the hard partonic part, 
and $\bar{T}$ is the hadronic matrix element:
\begin{eqnarray}
&&\bar{T}^{(I)}(x,x_1,x_2,k_T) \nonumber \\
&=&\int \frac{dy^-}{4\pi} \,
\frac{dy_1^-}{2\pi}\, 
\frac{dy_2^-}{2\pi}\, 
\frac{d^2y_T}{(2\pi)^2} \,    
e^{ixp^+y^-}\, e^{ix_1p^+(y_1^--y_2^-)}\, 
e^{ix_2p^+y_2^-}\, e^{ik_T \cdot y_T}\,
\nonumber \\
&\  & \times \,
\langle p_A |
A^+(y_{2}^{-},0_{T})\, \bar{\psi}_q(0)\, \gamma^+ \,
 \psi_q(y^{-}) \, A^+(y_{1}^{-},y_{1T}) | p_{A}\rangle \ .
\label{Tbari}
\end{eqnarray}
Because of the exponential factors in Eq.~(\ref{Tbari}), the position
space integration, $dy^-$'s, cannot give large dependence on the  
nuclear size unless the parton momentum fraction in one of 
the exponentials vanishes.  If the exponential vanishes, the corresponding
position space integration can be extended to the size of the whole 
nucleus.  Therefore, in order to get large nuclear enhancement
or jet broadening, we need to consider only Feynman diagrams 
that can provide poles which set parton momentum fractions on
the exponentials to be zero \cite{LQS1,LQS2}.  
Other diagrams that do not provide such poles will be suppressed 
by a large off-shell propagator and  hence are not leading contributions.
At the leading order in $\alpha_s$, only diagrams
shown in Fig.~\ref{fig1} have the necessary poles.
These diagrams contribute to the double scattering partonic
part $\bar{H}(x,x_1,x_2,k_T,x'p',p,q)$ in Eq.~(\ref{dyD}).

For the leading order diagrams shown in Fig.~\ref{fig1}, the 
corresponding partonic parts have two possible poles from the 
two propagators.  For the diagram shown in Fig.~\ref{fig1}a, 
the partonic part has the following general structure 
\begin{eqnarray}
\bar{H}^a
& \propto & 
\frac{1}{x_1-\frac{k_T^2}{x's}+i\epsilon}
\cdot \frac{1}{x_1-x_2-\frac{k_T^2}{x's}-i\epsilon} 
\nonumber \\
& \ & \times \delta (x+x_1-\frac{k_T^2}{x's}-\frac{Q^2}{x's}) \,
\delta (q_T^2-k_T^2) \ .
\label{dypole-a}
\end{eqnarray}
The two $\delta$-functions are from the phase space.
One of the $\delta$-functions  can be used 
to fix $dx$ integration in Eq.~(\ref{dyD}), and the two poles 
in Eq.~(\ref{dypole-a})  can be used to perform the contour 
integration for $dx_1 \, dx_2$.  We have 
\begin{eqnarray}
d\hat{\sigma}^D_a 
&\propto & (4\pi^2)\,\theta (y^--y_1^-) \,\theta (-y_2^-)\, 
 \delta (q_T^2-k_T^2)\, e^{i \, (\tau/x')\, p^+ y^-}\ ,
\label{dyIa}
\end{eqnarray}
with $\tau=Q^2/s$.  The two $\theta$-functions are the results of the
contour integration.

For the diagrams shown in 
Fig.~\ref{fig1}b and \ref{fig1}c, we have 
slightly different phase space factors and different poles. 
Similarly, after integrating over parton momentum fractions, 
$dx\, dx_1\, dx_2$, we have 
\begin{eqnarray} 
d\hat{\sigma}^D_b & \propto & 
(-4\pi^2)\, \theta (y_2^--y_1^-) \,
\theta (y^--y_2^-) \,\delta (q_T^2) e^{i \, (\tau/x')\, p^+ y^-}\ ,
\label{dyIb}
\end{eqnarray}
and
\begin{eqnarray}
d\hat{\sigma}^D_c & \propto &  (-4\pi^2)\, \theta (y_1^--y_2^-) \,
\theta  (-y_1^-) \,\delta (q_T^2)  e^{i \, (\tau/x')\, p^+ y^-}\ .
\label{dyc}
\end{eqnarray}

The  spinor trace gives the same
numerator for all three diagrams. Therefore, the total contribution to
transverse momentum broadening is proportional to 
$d\hat{\sigma}^D_a+d\hat{\sigma}^D_b+d\hat{\sigma}^D_c$, 
and this sum has the following feature
\begin{eqnarray}
\frac{d\sigma^D_{hA}}{dQ^2\,dq_T^2}
&\propto & d\hat{\sigma}^D_a+d\hat{\sigma}^D_b+d\hat{\sigma}^D_c
\nonumber \\
&\propto  &  \theta (y^--y_1^-) \, 
\theta (-y_2^-) \, \left[ \delta (q_T^2-k_T^2) - \delta (q_T^2) \right] 
\nonumber \\
&\ &+ \left[
\theta (y^--y_1^-) \,\theta (-y_2^-) -
\theta (y_2^--y_1^-)\, \theta (y^--y_2^-) \right.
\nonumber \\
&\ & \quad \left. -
\theta (y_1^--y_2^-)\, \theta (-y_1^-) \right]   \,
\delta (q_T^2) \ .
\label{Isum}
\end{eqnarray} 
It is clear from Eq.~(\ref{Isum}) that for the inclusive Drell-Yan cross 
section, $d\sigma/dQ^2$, the double scattering contribution, 
$d\sigma^D_{hA}/dQ^2$, does not have a large dependence on the nuclear size.
The integration over $dq_T^2$ eliminates the first term in Eq.~(\ref{Isum}),
while the second term is localized in space if $\tau/x'$ is not too 
small.  When the $\tau/x'$ is finite, and $p^+$ is large, 
exp$[i\,(\tau/x') p^+y^-]$  effectively
restricts $y^{-}\sim 1/((\tau/x') p^+) \rightarrow 0$.  
When $y^-\rightarrow 0$, the combination of 
the three pairs of $\theta$-functions in Eq.~(\ref{Isum}) vanishes,
\begin{eqnarray}
\frac{d\sigma^D_{hA}}{dQ^2} &\equiv &
\int dq_T^2 \left( \frac{d\sigma^D_{hA}}{dQ^2dq_T^2} \right)
\nonumber \\
&\propto &
\left[
\theta (y^--y_1^-) \,\theta (-y_2^-) -
\theta (y_2^--y_1^-)\, \theta (y^--y_2^-) -
\theta (y_1^--y_2^-)\, \theta (-y_1^-) \right]
\nonumber \\
& \rightarrow & 0 \quad\quad \mbox{as $y^-\rightarrow 0$} \ .
\label{totalD}
\end{eqnarray}
Physically, Eq.~(\ref{totalD}) says that 
all integrations of $y^-$'s are localized.
Actually, at the leading order, the term proportional to 
$\theta (y^--y_1^-) \,\theta (-y_2^-) -
\theta (y_2^--y_1^-)\, \theta (y^--y_2^-) -
\theta (y_1^--y_2^-)\, \theta (-y_1^-)$ is the eikonal contribution 
to make the normal twist-2 quark distribution
for single scattering gauge invariant.
Eq.~(\ref{totalD}) is a good example to demonstrate that 
the double scattering does not give a large nuclear size effect 
to the total inclusive cross section.

On the other hand, from Eq.~(\ref{Isum}),  
the double scattering contribution to the 
averaged transverse momentum square can give a large nuclear size effect
\begin{eqnarray}
\Delta \langle q_T^2 \rangle 
& \sim &
\int dq_T^2 \, q_T^2\, 
\left( \frac{d\sigma^D_{hA}}{dQ^2dq_T^2} \right)
\nonumber \\
&\propto &
\int dq_T^2 \, q_T^2\,
\left[ \delta (q_T^2-k_T^2) - \delta (q_T^2) \right] 
\nonumber \\
&\sim & k_T^2 \ .
\label{qt2Da}
\end{eqnarray}
Actually, $k_T^2$ in Eq.~(\ref{qt2Da}) needs to be integrated first. But 
Eq.~(\ref{qt2Da}) already demonstrates that 
$\Delta \langle q_T^2 \rangle$ is proportional 
$k_T^2$, which is the kick of the transverse momentum the parton received 
from the additional scattering. The bigger the nuclear size, the bigger 
the effective $k_T^2$. As shown below,  $\Delta \langle q_T^2 \rangle$ is 
proportional to the nuclear size.

After working out the algebra, we 
obtain the Drell-Yan transverse momentum broadening 
at the leading order in $\alpha_s$
\begin{equation}
\Delta \langle q_T^2 \rangle  
=\left(\frac{4\pi^2 \alpha_s}{3} \right)\cdot
\frac{\sum_{q} \, e_q^2\int dx' \, \phi_{\bar{q}/h}(x')\, 
T^{(I)}_{q/A}(\tau /x') /x'}
{\sum_{q}\, e_q^2 \int dx' \, \phi_{\bar{q}/h}(x')\, 
\phi_{q/A}(\tau /x) /x'} \ .
\label{dyqt2b}
\end{equation}
In Eq.~(\ref{dyqt2b}), $\phi_{q/A} (x)$ is the usual quark 
distribution inside a nucleus. $T_{q/A}^{(I)}(x)$
is the four-parton correlation function, and  is given by
\begin{eqnarray}
T_{q/A}^{(I)}(x) &=&
 \int \frac{dy^{-}}{2\pi}\, e^{ixp^{+}y^{-}}
 \int \frac{dy_1^{-}dy_{2}^{-}}{2\pi} \,
      \theta(y^{-}-y_{1}^{-})\,\theta(-y_{2}^{-}) \nonumber \\
&\ & \times \,
     \frac{1}{2}\,
     \langle p_{A}|F_{\alpha}^{\ +}(y_{2}^{-})\bar{\psi}_{q}(0)
                  \gamma^{+}\psi_{q}(y^{-})F^{+\alpha}(y_{1}^{-})
     |p_{A} \rangle \ .
\label{dyTq}
\end{eqnarray}
The superscript (I) represents the
initial state interaction, in order to distinguish from the 
similar four-parton correlation function defined in Eq.~(\ref{Ta}). 
In deriving Eq.~(\ref{dyqt2b}), we expend $\delta (q_T^2-k_T^2)$ 
at $k_T=0$, known as the collinear expansion, and keep only the first 
non-vanishing term which corresponds to 
the second order derivative term.
We use the factor $k_T^{\alpha} k_T^{\beta}$  
to convert the $k_T^{\alpha}A^+k_T^{\beta}A^+$ into field strength 
$F^{\alpha+}F^{\beta+}$ by partial integration. 
Here, we work in Feynman gauge. The terms associated with 
other components of $A^\rho$ 
are suppressed by $1/p^+$ compared to those with 
$A^+$, because of the requirement 
of the Lorentz boost invariance for the matrix elements \cite{LQS1}.

By comparing the operator definitions of these four-parton 
correlation functions and
the definitions of the normal twist-2 parton distributions,  
LQS proposed the following model  \cite{LQS1,LQS2}:
\begin{equation}
T_{f/A}(x)=\lambda^2 A^{4/3} \phi_{f/N}(x) \ ,
\label{TiM}
\end{equation}
where $\phi_{f/N}(x)$ with $f=q,\bar{q},g$ are the normal twist-2 parton 
distribution of a nucleon, and $\lambda$ is a free parameter 
to be fixed by experimental data. Applying this model, we obtain a very 
simple result for the  Drell-Yan transverse momentum
broadening:
\begin{equation}
\Delta \langle q_T^2 \rangle = 
\left(\frac{4\pi^2\alpha_s}{3} \right) \
\lambda^2\ A^{1/3} \ .
\label{dyqt2c}
\end{equation} 
Using Eq.~(\ref{dyqt2c}) and data from E772 and NA10 on nuclear 
enhancement of the average Drell-Yan transverse momentum,
$\Delta \langle q_T^2 \rangle$, we estimate 
the value of $\lambda^2$ as
\begin{equation}
\lambda^2_{\mbox{\tiny DY}} \approx 0.01 \mbox{GeV}^2 \ ,
\label{dylambda}
\end{equation}
which is at least a factor of five smaller than 
$\lambda^2_{\mbox{\tiny di-jet}} \approx 0.05- 0.1$ GeV$^2$,
previously estimated by LQS from momentum imbalance of the di-jet data 
\cite{LQS2}. Therefore, it is very important to use other observables, 
such as jet broadening in DIS, to further test the four-parton correlation
functions. 

\section{Jet broadening in Deeply Inelastic Scattering}

In this section, we derive the leading contribution to
jet broadening in DIS by using the same technique 
used to derive the Drell-Yan transverse momentum broadening, 
$\Delta \langle q_T^2 \rangle$,  
in last section.
Consider the jet production in  the deeply inelastic lepton-nucleus 
scattering, $e(k_1) + A(p) \rightarrow e(k_2) +jet(l) +X$. 
$k_1$ and $k_2$ are the four
momenta of the incoming and the outgoing leptons respectively,  
and $p$ is the momentum per nucleon for the nucleus with the atomic 
number $A$. With $l$ be the four-momentum for the jet,  
the averaged 
jet transverse momentum square is defined as
\begin{equation}
\langle l_T^2\rangle^{eA} =\left. \int dl_T^2 \cdot l_T^2 \cdot  
\frac{d\sigma_{eA}}{dx_BdQ^2dl_T^2} \, 
\right/ \frac{d\sigma_{eA}}{dx_BdQ^2} \ .
\label{lt2}
\end{equation}
where $x_B=Q^2/(2p\cdot q)$, and 
$q=k_1-k_2$ is the momentum of the virtual photon, and $Q^2=-q^2$. 
The jet transverse momentum $l_T$ depends on our choice of the frame. 
We choose the Breit frame in the following calculation. 
Similar to the Drell-Yan transverse momentum spectrum, 
$d\sigma/dQ^2 dq_T^2$, the jet transverse 
momentum spectrum, $d\sigma/dx_BdQ^2dl_T^2$, is sensitive to 
the $A^{1/3}$ type nuclear size effect due to the multiple scattering. 
On the other hand, the inclusive DIS cross section 
$d\sigma/dx_BdQ^2 = \int d l_T^2\, d\sigma/dx_BdQ^2dl_T^2$ 
does not have the $A^{1/3}$ power enhancement.  Instead, it has 
a much weaker A-dependence, such as the EMC effect and the nuclear 
shadowing.  
To separate the multiple scattering contribution from 
the single scattering, we define the jet broadening as 
\begin{equation}
\Delta \langle l_T^2\rangle  \equiv \langle l_T^2\rangle ^{eA} -
\langle l_T^2 \rangle ^{eN} \ ,
\label{broaden} 
\end{equation}
Keeping only the contribution from the double scattering, 
similar to Eq.~(\ref{dydqt2a}), we have 
\begin{equation}
\Delta \langle l_T^2\rangle
\approx
\left. \int dl_T^2 \cdot l_T^2 \cdot 
\frac{d\sigma^{D}_{eA}}{dx_BdQ^2dl_T^2} \, \right/ 
\frac{d\sigma_{eA}}{dx_BdQ^2} \ .
\label{dlt2d}
\end{equation} 

\begin{figure}[hbt]
\begin{minipage}[t]{2.0in}
\psfig{figure=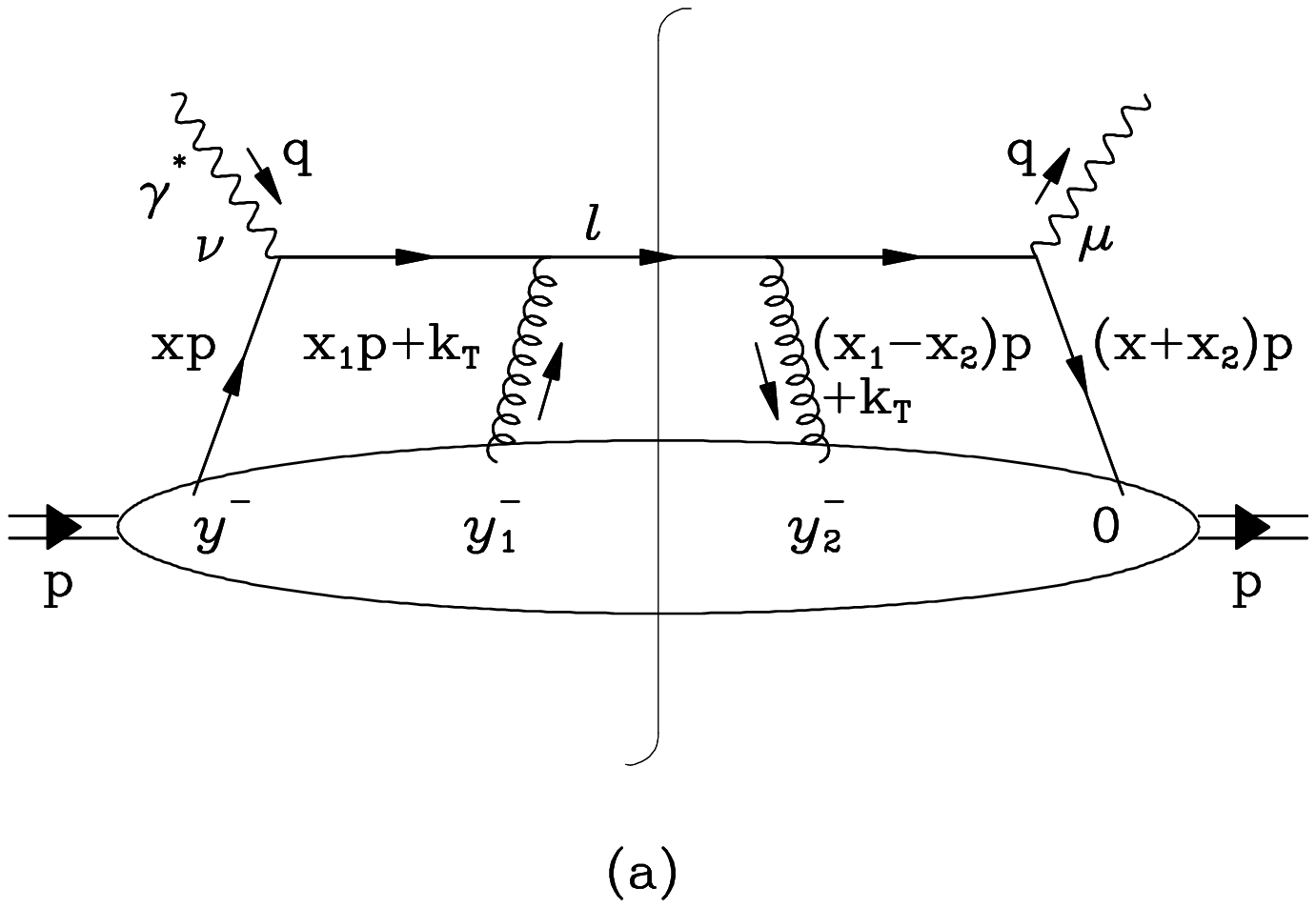,width=1.0in}
\end{minipage}
\hfill
\begin{minipage}[t]{2.0in}
\psfig{figure=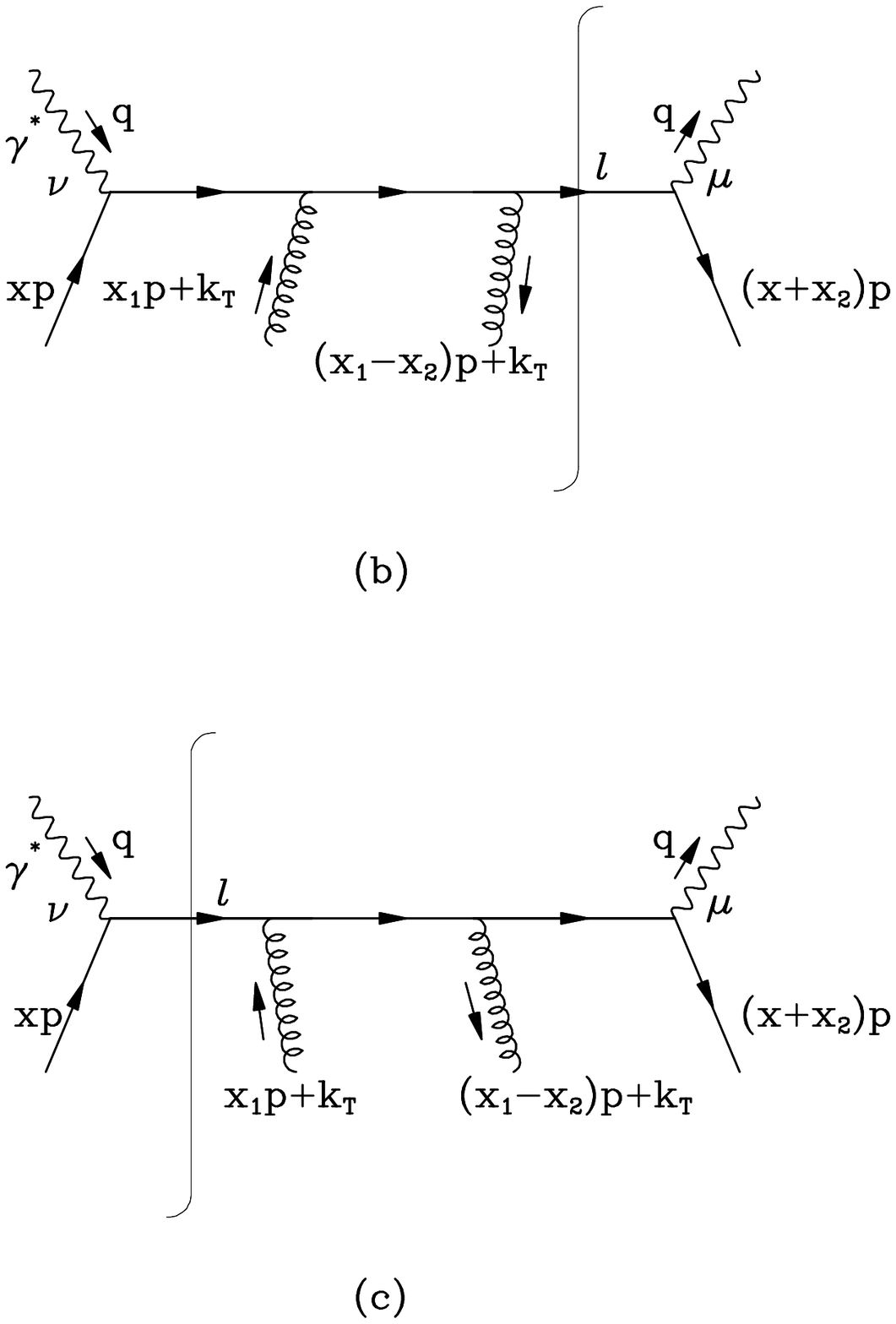,width=1.0in}
\end{minipage}
\caption{Lowest order double scattering contribution to jet broadening: 
(a) symmetric diagram; (b) and (c): interference diagrams.}
\label{fig3}
\end{figure}

The leading order double scattering diagrams contributing to 
jet broadening are given in Fig.~\ref{fig3}. 
The double scattering contribution  can be expressed as
\begin{equation}
d\sigma^D \propto 
\int dx\, dx_1\, dx_2 \int d k_T^2 \, 
\bar{T}^{(F)}(x,x_1,x_2,k_T) 
\bar{H}_{\mu\nu}(x,x_1,x_2,k_T,p,q,l) \ ,
\label{W1}
\end{equation}
with the matrix element 
\begin{eqnarray}
&& \bar{T}^{(F)}(x,x_1,x_2,k_T)  \nonumber \\
&=&
\int \frac{dy^-}{4\pi}\, \frac{dy_1^-}{2\pi}\, 
\frac{dy_2^-}{2\pi} \,
\frac{d^2y_T}{(2\pi)^2} \,    
 \times e^{ixp^+y^-}\, e^{ix_1p^+(y_1^--y_2^-)}\, 
e^{ix_2p^+y_2^-}\, e^{ik_T \cdot y_T}
\nonumber \\
&\  &\, \times \,\langle p_A |
 \bar{\psi}_q(0)\, \gamma^+ \,A^+(y_{2}^{-},0_{T})\,
          A^+(y_{1}^{-},y_{1T})\, \psi_q(y^{-}) | p_{A}\rangle \ ,
\label{TbarF}
\end{eqnarray}
where superscript ($F$) indicates the final-state double scattering.
The matrix element $\bar{T}^{(F)}$ is equal to the  
$\bar{T}^{(I)}$ in Eq.~(\ref{Tbari}) if we commute the gluon fields 
with the quark fields \cite{QS,jaffe}.
In Eq.~(\ref{W1}), $\bar{H}_{\mu\nu}$ is the corresponding partonic 
part. Here, we work in Feynman gauge.
For the diagram shown in Fig.\ref{fig3}a, the partonic part  has the
following structure
\begin{eqnarray}
\bar{H}^a_{\mu\nu} 
& \propto & 
\delta (x+x_1-x_B -\frac{k_T^2-2q\cdot k_T}{2p\cdot q}) \cdot
\delta (l_T^2 -k_T^2) \, dl_T^2  
\nonumber \\
&\ & \times 
\frac{1}{x-x_B+i\epsilon} \cdot 
\frac{1}{x+x_2-x_B-i\epsilon} \ .
\label{Hbar}
\end{eqnarray}
In Eq.~(\ref{Hbar}), the $\delta$-function is from the phase space, and 
the poles are from the propagators.

Following the derivation of  $\Delta \langle q_T^2 \rangle$ in last section, 
first, 
we carry out the integrations of the parton momentum fractions by
using the $\delta$-function and two poles in Eq.~(\ref{Hbar}),
\begin{eqnarray}
d\sigma^a
&\propto & 
(2\pi)^2\, \theta (y_1^--y^-) \,\theta (y_2^-)\, 
\delta(l_T^2-k_T^2) \ .
\label{intHa}
\end{eqnarray}
Similarly, we have the corresponding integrations for the interference 
diagram in Fig.~\ref{fig3}b
\begin{eqnarray}
d\sigma^b
& \propto & - (2\pi)^2\, \theta (y_2^--y_1^-) \,\theta (y_1^--y^-)\, 
 \delta(l_T^2)  \ ,
\label{intHb}
\end{eqnarray}
and for the diagram in Fig.~\ref{fig3}c
\begin{eqnarray}
d\sigma^c 
& \propto & -(2\pi)^2\, \theta (y_2^-) \,\theta (y_1^--y_2^-)\, 
 \delta(l_T^2)  \ .
\label{intHc}
\end{eqnarray}
Combining Eqs.~(\ref{intHa}), (\ref{intHb}) and (\ref{intHc}), we 
again have the same structure as that in Eq.~(\ref{Isum}).  Therefore, 
like in the Drell-Yan case, we conclude that the 
double scattering does not result 
into any large nuclear size dependence in the inclusive DIS cross section. 
For the jet broadening defined in Eq.~(\ref{dlt2d}), we drop 
the term proportional to $\theta (y_1^--y^-) \,\theta (y_2^-) -
\theta (y_2^--y_1^-)\, \theta (y_1^--y^-) -
\theta (y_1^--y_2^-)\, \theta (y_2^-)$, which is 
localized as the single scattering.  After carrying out
the algebra, we derive jet broadening in DIS as
\begin{equation}
\Delta \langle l_T^2 \rangle 
= \left(\frac{4\pi^2 \alpha_s}{3} \right)
\cdot \frac{\sum_{q}\, e_q^2\, T_{q/A}(x_B)}
           {\sum_{q}\, e_q^2\, \phi_{q/A}(x_B)} \ ,
\label{lt2T}
\end{equation}
where $\sum_q$ sums over all quark and antiquark flavors.
In Eq.~(\ref{lt2T}), $\phi_{q/A}(x)$ is the normal twist-2 quark 
distribution inside a nucleus, and the four-parton correlation
function, $T_{q/A}(x_B)$ is defined as \cite{LQS2}
\begin{eqnarray}
T_{q/A}(x) &=&
\int \frac{dy^-}{2\pi}\, e^{ixp^+y^-}\
\frac{dy_1^-dy_2^-}{2\pi}  
\theta (y_1^--y^-) \,\theta (y_2^-)\, 
\nonumber \\
&\ & \times \, \frac{1}{2}\,
\langle p_A |
 \bar{\psi}_q(0)\, \gamma^+ \,F_{\sigma}^{\ +}(y_{2}^{-})\,
          F^{+\sigma}(y_{1}^{-})\, \psi_q(y^{-}) | p_{A}\rangle \ .
\label{Ta}
\end{eqnarray}
With the model given in Eq.~(\ref{TiM}), we can simplify 
Eq.~(\ref{lt2T}):
\begin{equation}
\Delta \langle l_T^2 \rangle = 
\left(\frac{4\pi^2\alpha_s}{3} \right) \
\lambda^2\ A^{1/3} \ .
\label{dlt2}
\end{equation}
From Eq.~(\ref{lt2T}), we see that jet broadening in DIS 
is directly proportional to the four-parton (twist-4) correlation 
function, defined in Eq.~(\ref{Ta}). It is the same four-parton
correlation function appeared in dijet momentum imbalance \cite{LQS2}.  
$T_{q/A}(x)$  and $T^{(I)}_{q/A}(x)$, which are 
defined in Eqs.~(\ref{Ta}) and (\ref{dyTq}), 
respectively, are same if the phase space integral are symmetric, because
field operators commute on the light-cone \cite{QS,jaffe}. 
Therefore we conclude that jet broadening in DIS, transverse 
momentum broadening for the Drell-Yan pairs, and dijet momentum 
imbalance depend on the same four-parton correlation function.


\section{Discussion and conclusions}

From the simple expression in Eq.~(\ref{dlt2}), we conclude
that at the leading order, jet broadening in DIS has a 
strong scaling property, that it does not depend on beam energy, 
$Q^2$ and $x_B$.  This scaling property of jet broadening in DIS is
a direct consequence of  LQS model for four-parton correlation 
functions, given in Eq.~(\ref{TiM}). However, when $x_B \leq 0.1$, 
$y^- \sim 1/(x_Bp^+) $ is no longer localized inside an individual 
nucleon \cite{strikman,MQ}. Therefore, terms proportional to 
 $\theta (y_1^--y^-) \,\theta (y_2^-) -
\theta (y_2^--y_1^-)\, \theta (y_1^--y^-) -
\theta (y_1^--y_2^-)\, \theta (y_2^-)$, are no longer localized  
 and need to be kept for the jet broadening calculation if $x_B$ is 
small.  Consequently, the $x_B$-scaling of jet
broadening in DIS needs to be modified in small $x_B$ region. 
In addition, $Q^2$-dependence may be modified because 
the four-parton correlation function
$T_{q/A}(x)$ and the normal quark distribution $\phi_{q/A}(x)$
can have different scaling violation. In principle, 
 all dependence or whole conclusion could be modified due
to possible different high order corrections.  Nevertheless, we
believe that experimental measurements of jet broadening in DIS 
can provide valuable information on the strength of multi-parton 
correlations and the dynamics of the multiple scattering.
 
Similarly, from Eq.~(\ref{dyqt2c}), we can also conclude that 
the Drell-Yan transverse momentum broadening, 
$\Delta \langle q_T^2 \rangle$, has a  
small dependence on beam energy and $Q^2$ of the lepton pair.
Data from Fermilab E772 \cite{E772} and CERN NA10 \cite{NA10}
demonstrate weak energy dependence.  It signals that the simple model 
by LQS for four-parton 
correlation functions is reasonable and the leading 
order calculation given here are useful. At the same time, the 
observed energy dependence indicates that the high order corrections to 
$\Delta \langle q_T^2 \rangle$ can not be ignored \cite{Guo,GQS}.

In addition, Eqs.~(\ref{dlt2}) and (\ref{dyqt2c})
tell us that at the leading 
order, jet broadening in DIS and nuclear enhancement of average 
Drell-Yan transverse momentum, $\Delta \langle q_T^2 \rangle$,  
have  the same  magnitude, if the averaged 
initial-state gluon interactions is equal to the corresponding 
final-state gluon interactions, i.e., 
$T_{q/A}(x)=T_{q/A}^{(I)}(x)$.  

From Eq.~(\ref{lt2T}), we see that the jet broadening 
is directly proportional to the four-parton correlation 
function $T_{q/A}(x_B)$. Information on $x_B$-dependence of the 
jet broadening can provide a first ever direct measurement of
the functional form of four-parton correlation functions 
$T_{q/A}(x_B,A)$. In addition, 
by examing the scaling property of jet broadening, we can 
directly  verify LQS model of four-parton correlation functions, given 
in Eq.~(\ref{TiM}).
Future experiments at HERA with a 
heavy ion beam  should be able to provide much more 
information on dynamics of parton correlations.

In summary, we have derived analytic expressions for jet broadening 
in DIS and the Drell-Yan transverse momentum broadening in terms 
of universal four-parton correlation functions. 
Because the dijet data (pure final-state multiple scattering)
and the Drell-Yan data (pure initial-state multiple scattering) 
favor two different sizes of the four-parton correlation function, 
measurement of jet broadening in DIS 
(pure final-state multiple scattering) will provide a critical test 
of QCD dynamics of the multiple scattering. Since at the leading 
order, the Drell-Yan transverse momentum broadening,
$  \Delta \langle q_T^2 \rangle _{\mbox{\tiny DY}}$,  
and jet broadening in DIS, 
$\Delta \langle l_T^2 \rangle _{\mbox{\tiny DIS}}$, 
 are independent of the four-gluon correlation 
function $T_{g/A}$ \cite{LQS2}, measurement of 
$\Delta \langle q_T^2 \rangle _{\mbox{\tiny DY}}$   
and $  \Delta \langle l_T^2 \rangle _{\mbox{\tiny DIS}}$ 
provide a direct comparison between the initial-state multiple 
scattering  and the final-state multiple scattering. 

\section*{Acknowledgments}
This work was partially supported by the U.S. Department 
of Energy under contract No. DE-FG02-93ER40764.

\end{document}